\documentclass[aps,twocolumn,prl,showpacs]{revtex4-1}

\usepackage{bm,graphicx}
\begin{document}

\title{Amplitude Fluctuations Driven by the Density of Electron Pairs within Nanosize Granular Structuters inside Strongly 
        Disordered Superconductors: Evidence for a Shell-Like Effect}
\author{Sanjib Ghosh and Sudhansu S. Mandal}
\affiliation{Department of Theoretical Physics, Indian Association for the Cultivation of Science,
           Kolkata 700 032, India}

\date{\today}

\begin{abstract}
 Motivated by the recent observation of the shell effect in a nanoscale pure superconductor by Bose {\em et al} [Nat. Mat. {\bf 9}, 550 (2010)],
we explore the possible shell-like effect in a strongly disordered superconductor as it is known to produce nanosize
superconducting puddles (SPs). We find a remarkable change in the texture of the pairing amplitudes that is responsible for forming the SP,
 upon monotonic tuning of the average electron density, $\langle n \rangle$, and keeping the disorder landscape unaltered. 
Both the spatially averaged pairing amplitude and 
the quasiparticle excitation gap oscillate with $\langle n \rangle$. This oscillation is due to a rapid change in the low-lying quasiparticle
energy spectra and thereby a change in the shapes and positions of the SPs.
We establish a correlation between the formation of SPs and the shell-like effect.
The experimental consequences of our theory are also discussed.     

\end{abstract}
\maketitle

A finite-size correction to the Bardeen-Cooper-Schrieffer theory \cite{BCS} of finite-size superconducting metallic
grains predicts \cite{Altshuler,Ovchinnikov,Olofsson} a large change in the energy gap due to a small change in electron number.
 The pairing amplitude (PA)
oscillates with the change in mean-level spacing which may be tuned by changing either the particle number or the
size and shape of the grains. This fluctuation arises due to a rapid change in the spectral density at low
energies. This phenomenon is known as the shell effect in small-size superconductors and was recently observed
by Bose {\em et al} \cite{Bose} in pure superconducting nanoparticles.

The superconductor-to-insulator transition (SIT) \cite{Goldman,Review1} with increasing disorder in thin films is an 
archetypal example for studying the competition between interaction and disorder. While the
attractive interaction between electrons is responsible for the formation of cooper pairs,
which condense into a macroscopic quantum state \cite{BCS} called superconductivity, the disorder
localizes the electronic states \cite{Anderson_loc}. The concomitance of these two contrasting quantum
states in the presence of both attractive interaction and disorder leads to several fascinating quantum
effects such as inhomogeneity and the formation of superconduting puddles (SPs) \cite{Ghosal,Dubi,Sacepe,Goldman3} in PA, 
the presence of a pseudogapped
\cite{Baturina2,Pratap1} phase where the long-range order of superconductivity diminishes although the quasiparticle gap remains open \cite{Ghosal},
and fractal superconductivity \cite{Feigelman,Ioffe} in which the correlations of certain  functions become fractal
in nature before they transform into completely localized states.
In this Letter, we show that the shell-like effect in disordered superconductors occurs because the 
systems mimic the collection of nanosize superconductors in the form of SPs.


Although the superconductors remain homogeneous \cite{Anderson,AG,Ma,Kapitulnik} in the low-disorder regime, inhomogeneity \cite{Ghosal,Sacepe}
in the PA develops with an
increase in the strength of disorder. In a moderate range of disorder, SPs with larger PA separated by insulating regions with
 vanishingly small PA are formed \cite{Ghosal}. The phase separation between the SPs and insulating puddles transforms  either of these
phases into the global phase. The SPs connected by Josephson tunneling give rise
to a global superconductor \cite{Fisher}, whereas the large phase fluctuations (PFs) between the SPs at high disorder drive \cite{Nandini}
the system into an insulating state.
Even in the insulating state, puddles with nonzero PA exist, and one finds a pseudogap \cite{Ghosal} in the spectral function. 
Since the typical length scales of these nanoscale puddles become much less than the system size, these SPs can resemble small size
superconductors.  
We thus focus here on the spatial variation of the PA as a function of mean electron density, $ \langle n \rangle$,
 with a fixed landscape of strongly disordered potential.
The average densities can be tuned without altering the disorder landscape by applying electrostatic gates \cite{Goldman2} to a thin film superconductor.

Our calculations below demonstrate that the texture of the PA forming the SPs at strong disorder, especially in the insulating side of the SIT,
undergoes a huge change with little change in $\langle n \rangle$, in spite of the unaltered landscape of disorder. This fact is not {\em a priori}
known since the little change in $\langle n \rangle$ does not change much of the local effective chemical potential and thereby
only a little of the local occupation number density. We find that as the bulk chemical potential changes with $\langle n \rangle$, the 
quasiparticle energy spectra (and the corresponding eigenstates) around the chemical potential change. The low-lying quasiparticle
eigenstates create \cite{Ghosal} the regions of higher the PA. Due to these reasons, the texture of PA and thus the shapes and positions of the SPs keep on changing
with $\langle n \rangle$. Moreover, the excitation gap and the spatially averaged PA, $\bar{\Delta}$, oscillate with $\langle n\rangle $. 
This shell-like effect in disordered superconductors occurs when the mean-level spacing of the low-lying quasiparticle eigenstates becomes comparable 
to the disorder-averaged $\bar{\Delta}$.


We employ the method of the self-consistent solution of the Bogoliubov-de Gennes (BDG) equation \cite{DeGennes}, which has
been described in detail by others \cite{Ghosal,Dubi}. In brief , 
we study the BDG equation  for a disordered superconductor at a site $i$ in a square lattice,
\begin{equation}
 \left[ \begin{array}{rr}
         {\cal H}_0  & \Delta_i\\
         \Delta_i & -{\cal H}_0 
        \end{array}
    \right] \left(  \begin{array}{c}
                     u_m^i \\ v_m^i
                    \end{array}  \right) = E_m  \left(  \begin{array}{c}
                     u_m^i \\ v_m^i
                    \end{array}  \right)
\label{BDG}
\end{equation}
with eigenvalues $E_m$, BDG amplitudes $u_m^i$ and $v_m^i$, pair-amplitude $\Delta_i$, 
${\cal H}_0u_m^i (v_m^i) = -t\sum_\delta u_m^{i+\delta} (v_m^{i+\delta}) +
 [V_i - \tilde{\mu}_i]u_m^i (v_m^i)$, where $\delta = \pm \hat{x}, \pm \hat{y}$,
local chemical potential $\tilde{\mu}_i = \mu + U n_i /2$ 
renormalized due to attractive on-site interaction $-U$, leading to $s$-wave superconductivity, hopping energy $t$, chemical
potential $\mu$ determined with the fixed average density $\langle n\rangle <1$, 
local density $n_i = 2\sum_m (v_m^i)^2$, local
pair-amplitude $\Delta_i = U \sum_m u_m^iv_m^i$, and the random potential $V_i$ at each site 
drawn from the uniform distribution in the range $[-V,\, V]$. Henceforth, all the energies are in the unit of nearest
neighbor hopping energy and all the length scales are in the unit of lattice constant. The BDG equation
(\ref{BDG}) is solved with periodic boundary conditions in a square lattice with $N$ sites. 
We determine $\Delta_i$ and $n_i$ for a fixed $\langle n\rangle=(1/N)\sum_i n_i$ and a chosen
random distribution of $V_i$ by iteratively solving Eq.~(\ref{BDG}) and 
above equations of $\Delta_i$ and $n_i$ until the self-consistency is reached.

We have made our calculations for a range of parameters: $1 \leq U \leq 4$, $ 1.5 \leq V \leq 4$ on square lattices
of size $N= L \times L$ with $L=24$, 32, and 40. We present \cite{note2} the results below for $U=1.5$, $V=2$, and $L=32$; similar
results are obtained for other parameters as well. We have checked that the solutions obtained in the  self-consistent
method are independent of initial guesses; the results related to self-consistency in Ref.~\onlinecite{Ghosal} are also 
reproduced by our calculation.

\begin{figure}[h] 
\centering
\includegraphics[scale=0.9]{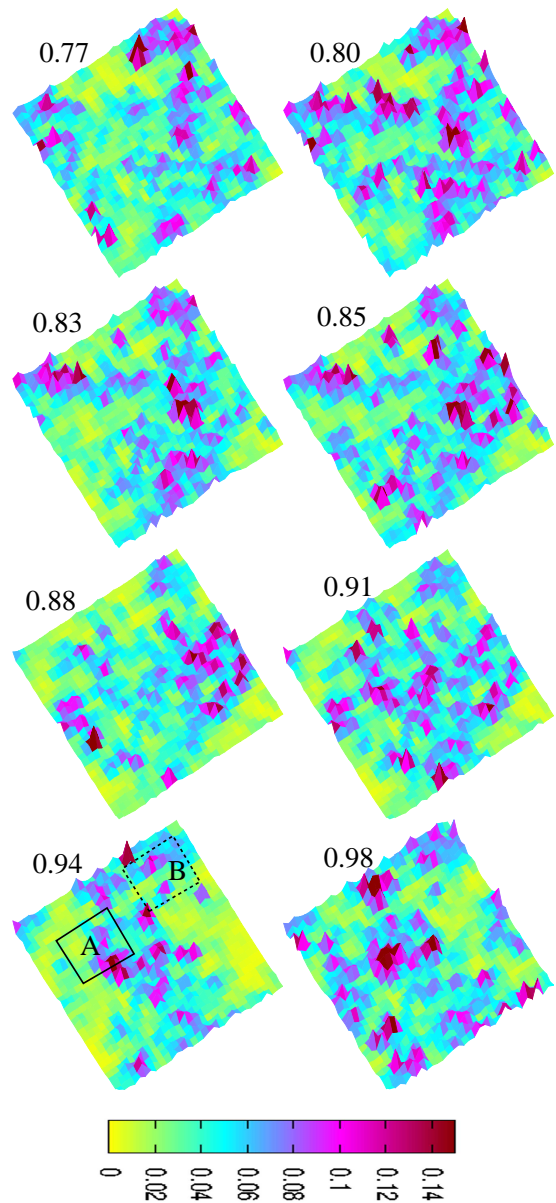}
 \caption{(color online) Topographic plot of pair-amplitudes at different sites in a $32\times 32$ square lattice (shown as rhomboidal plane) with $U=1.5$ 
at different values of $\langle n \rangle$, given adjacent to
each curve for a fixed realization of random disorder with $V=2$. The shapes and positions of the superconducting islands change 
with $\langle n \rangle$. In the left bottommost panel, two squares of size $10\times  10$ that will be used in Fig.~3 are marked
as A and B. }
\label{Fig1}
\end{figure}

Figure \ref{Fig1} represents a systematic study of the spatial variation of $\Delta_i$ for a fixed realization of disorder but with
different values of mean carrier densities. Several features of this comparative study are noteworthy:
(i) the SPs separated by insulating regions are formed, as expected,  but the shapes and positions of the SPs change 
rapidly with $\langle n \rangle$, (ii) 
the system passes through a relatively less inhomogeneous structure between two entirely different types of 
strongly inhomogeneous structures if $\langle n \rangle$
is tuned, {\it e.g.}, a less inhomogeneous structure at $\langle n \rangle = 0.91$ between two highly inhomogeneous structures at 
$\langle n \rangle = 0.88$ and 0.94 (Fig.~1), (iii) a site that may lie in the insulating region for a certain density, it becomes part of a
superconducting island at certain other densities, and (iv) the change in the texture of the PA is nonmonotonic with $\langle n \rangle$.

\begin{figure}[h]
\centering
\includegraphics[scale=0.5]{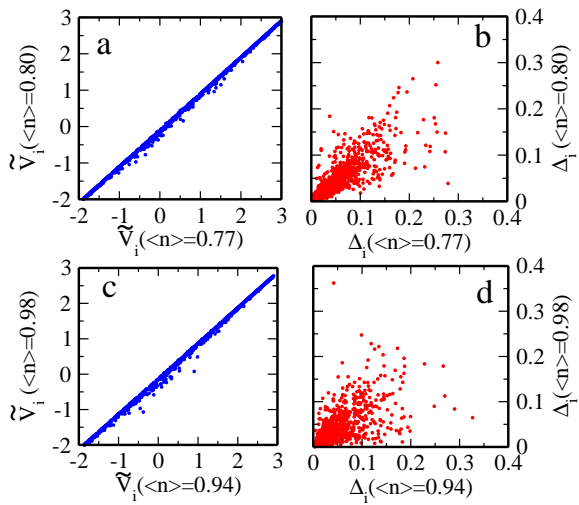}
\caption{(color online) Comparison of (a, c) effective disorder potential, $\tilde{V}_i$, and (b, d) pair-amplitude, $\Delta_i$, at two sets of
two different
average densities that are closed to each other: (a, b) $\langle n \rangle = 0.77$ and 0.80; (c,d) $\langle n \rangle = 0.94$ and 0.98.
Here each point corresponds to the same site $i$ for  the quantities in both the horizontal and vertical axes.} 
 \label{Fig2}
  \end{figure}

\begin{figure}[h]
\centering
\includegraphics[scale=0.45]{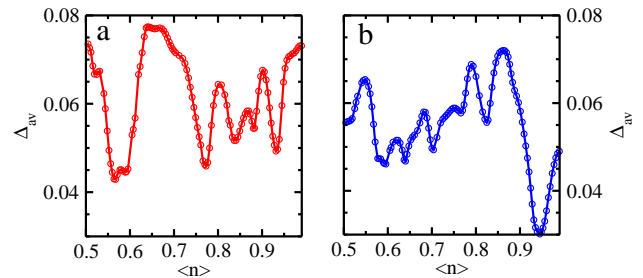}
\caption{(color online) Variation of average pair-amplitude, $\Delta_{{\rm av}}$, 
 taken in two small regions (marked in Fig.~\ref{Fig1}) of sizes $10 \times 10$ in a $32 \times 32$ square
lattice, with  $\langle n \rangle$. $V_i$ is kept the same for all densities. Left (right) panel corresponds to square A (B) 
marked in Fig.~\ref{Fig1}. \label{Fig3} }
  \end{figure}

It was argued before \cite{Ghosal} that the strong fluctuation
in $\Delta_i$ is correlated with the effective disorder potential $ \tilde{V}_i=V_i-\tilde{\mu}_i$: $\Delta_i$ is large at 
those sites where $\vert \tilde{V}_i \vert$ is small and vice versa. In this respect, any small change \cite{note} in $\langle n \rangle$
will only have little effect on $\Delta_i$, especially for the sites at which $\Delta_i$ are small; there will be minor readjustment of the 
values of $\Delta_i$ within a superconducting puddle. Therefore, no change in the positions of SPs is {\it a priori} expected.
To examine the validity of this fact, we compare $\Delta_i$ and $\tilde{V}_i$ 
 between $\langle n \rangle = 0.77$  and $0.80$ and also between $0.94$ and $0.98$ (Fig.~\ref{Fig2}).
Although $\tilde{V}_i$ remains almost unchanged (Figs.~\ref{Fig2}a, \ref{Fig2}c) at all sites (as the data follow a straight line with unit slope) for not
so different values of $\langle n \rangle$, $\Delta_i$ changes substantially. 
While $\Delta_i$ in a given site $i$ is small at some density,
it may be large for some other densities that are not very different (Figs.~\ref{Fig2}b, \ref{Fig2}d) from the former. 
 This proves that the fluctuation in $\Delta_i$ is not strongly
correlated with $\tilde{V}_i$. We will show below that the fluctuation in $\Delta_i$ and the formation of islands with larger PA is 
correlated with the shell-like effect that predisposes in favor of forming SPs, which in turn reproduces shell effect.
To gain further insight, we choose two small
regions of size $10\times 10$ each (shown as two square regions A and B in Fig.~\ref{Fig1}) in a $32\times 32 $ lattice and calculate average PA,
$\Delta_{{\rm av}} = (1/M)\sum_{i=1}^M \Delta_i$ with $M$ being the total number of sites in each region, for these regions with
different values of $\langle n \rangle$. Figure \ref{Fig3} shows rapid oscillation \cite{note1} 
of $\Delta_{{\rm av}}$ with $\langle n \rangle$; the deviation
of $\Delta_{{\rm av}}$ occurs up to about $50\%$ of its maximum value. 
This rules out the possibility of forming SPs at a given region of space at all densities.
The lower (higher) the values of $\Delta_{{\rm av}}$ in the selected region, the more susceptible the region becomes to becoming part of an insulating 
(superconducting) region.

\begin{figure}[h]
\centering
\includegraphics[scale=0.35]{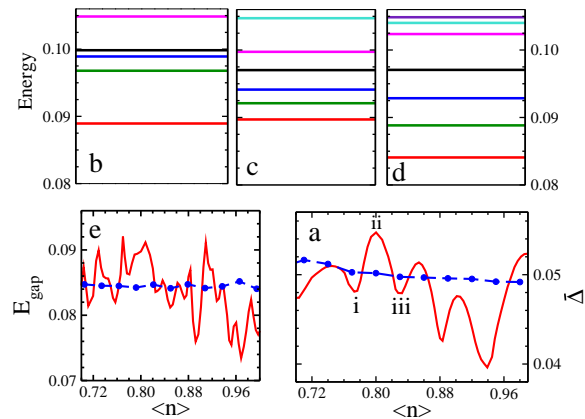}
\caption{(color online) Oscillation of (a) mean pair-amplitude , $\bar{\Delta}$,  in a $32 \times 32$ lattice for $U=1.5$, $V=2$, and a 
given realization of disorder,
  with $\langle n \rangle$. Dashed line with points represents disorder average $\bar{\Delta}$ at $V=2$. 
Low-lying quasiparticle eigenenergies (b, c, d) at three chosen densities marked by respective points (i), (ii),
and (iii) in (a) corresponding to three different extrema in $\bar{\Delta}$. (e) Variation of gap energy, $E_{{\rm \small{gap}}}$, with $\langle 
n \rangle$. It also oscillates with $\langle n \rangle$. The dashed line represents the disorder-averaged energy gap at $V=2$.
Disorder averages are taken with 12 realizations of disorder.  \label{Fig4}   }
  \end{figure}

\begin{figure}[h]
\centering
\includegraphics[scale=0.5]{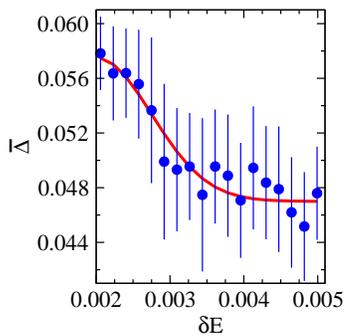}
\caption{(color online)  The plot of the average value of $\bar{\Delta}$ in the range of mean level spacing between $\delta E$ and $\delta E + 0.0002$ against
$\delta E$, which has been calculated for the lowest ten quasiparticle eigenenergies. $\bar{\Delta}$ and $\delta E$ have been calculated for 
$U=1.5$, $V=2$, 126 values of $\langle n \rangle$ between 0.7 and 1.0, and ten realizations of disorder. The vertical line about the average value of
$\bar{\Delta}$ is its standard deviation. 
 The solid line is a guide to the eye to show the decrease of $\bar{\Delta}$ with the increase of $\delta E$. \label{Fig5} }
  \end{figure}

The change in spatial fluctuation of $\Delta_i$ with $\langle n \rangle$ is also responsible  for changing the average PA of the system. As we find
here, the mean PA, $\bar{\Delta}= (1/N)\sum_i\Delta_i$, of the system oscillates with $\langle n \rangle$ [Fig.~\ref{Fig4}a] for a given disorder landscape.
However, the disorder-averaged $\bar{\Delta}$ is almost independent of $\langle n \rangle$ [Fig.~\ref{Fig4}a] \cite{Kumar}.
The lower (higher) values of $\bar{\Delta}$ correspond to higher (lesser) inhomogeneity in the system.  
This oscillation occurs
due to rapid change in the low-lying  quasiparticle energy spectra, analogous to the shell effect in superconducting nanoparticles.
 Figures \ref{Fig4}b, \ref{Fig4}c, and \ref{Fig4}d show low-energy quasiparticle eigenstates at three densities
that correspond to three consecutive extrema in Fig.~\ref{Fig4}a. There are two important characteristics to be noted from these quasiparticle spectra:
(i) the quasiparticle gap, $E_{{\rm gap}}$,  and the mean level spacing, $\delta E$,  for low-lying states change
with $\langle n \rangle$; and (ii) $E_{{\rm gap}}$ oscillates with $\langle n \rangle$ (Fig.~\ref{Fig4}e) for a given disorder landscape, although its disorder-averaged value is independent of $\langle n \rangle$. 
The oscillations in $E_g$ and $\bar{\Delta}$ are found to be up to 25$\%$ of their respective maximum values.
We calculate $\bar{\Delta}$ for ten sets of disorder realization and at $126$ densities between $\langle n \rangle = 0.7$ and $1.0$,
determine mean level spacing, $\delta E$, for the lowest ten eigenenergies in each of these cases, and plot average value of $\bar{\Delta}$
in the range $\delta E$ and $\delta E + 0.0002$ against $\delta E$ in Fig.~\ref{Fig5}.
Although $\bar{\Delta}$ oscillates and quasiparticle energy spectra change with $\langle n \rangle$, 
 the former seems to be monotonically correlated with $\delta E$. 

In a finite-size superconducting grain, the level spacing depends on the size of the grain. When the average level spacing becomes
comparable to the bulk gap, the shell effect is observed \cite{Bose}. The PA depends on the number of available quasiparticle 
states in a narrow window of Debye energy about the Fermi energy. Since the number of states within this window fluctuates with the 
shifting of Fermi energy or, equivalently, with the change in electron density, the fluctuation in PA occurs.
In the present system of the disordered superconductor, as Ghosal et al \cite{Ghosal} pointed out, low-lying quasiparticle eigenstates lie in the 
SPs and thus the quasiparticle gap remains open even at high disorder. These SPs are of nanoscale size and behave as  small-size superconductors
that become sensitive to shell-like effect when the mean level spacing for low-lying quasiparticle states becomes comparable 
(within one order less in magnitude) to the 
average PA, $\bar{\Delta}$. In the strongly disordered superconductor, mean level spacing $\sim (\pi/\xi_{{\rm loc}})^2$ (where $\xi_{{\rm loc}}$
is the localization length) increases \cite{Ma} 
with the increase of the strength of disorder $V$, and disorder-averaged $\bar{\Delta}$ decreases with the decrease of $U$.
Therefore the shell-like effect will be prominent on lowering \cite{note3}
 $U$ for a moderate $V$, and on increasing $V$ for a moderate $U$.
We have shown here that the substantial change in low-lying quasiparticle
eigenenergies, $E_m$, and the corresponding eigenfunctions $(u_m^i,\, v_m^i)$ occur by tuning $\langle n \rangle$. 
This causes to change in the positions of the superconducting and insulating islands as shown in Fig.~\ref{Fig1}.

 The systematic control of changing electron density without changing the landscape
of disorder in the thin-film superconductor has already been demonstrated \cite{Goldman2}. The scanning tunneling microscopic  measurements \cite{Sacepe}
in such an arrangement would 
directly show the change in position and shape of the superconducting islands with density. In the same experiment, the average gap
over a certain region of the system should oscillate with density. 
This effect will also 
occur in the pseudogapped insulating phase since the inhomogeneity in PA is already reported \cite{Goldman3} in the insulating phase. 
Although the SIT can only be considered when phase fluctuations are included on top of the mean field studied here, one may naively consider
the lesser (higher) inhomogeneous structure as superconducting (insulating) phase. Therefore, the possibility of a reentrant phase of superconductivity
observable in resistivity measurements may not be ruled out, upon tuning $\langle n \rangle$ at large disorder.

There have been various studies such as the self-consistent solution of BDG equations \cite{Ghosal,Kamlapure,Dubi2,Seibold}, 
classical Monte Carlo calculations at finite temperatures \cite{Dubi,Erez}, and 
quantum Monte Carlo calculations \cite{Nandini} performed using negative-$U$ Hubbard interaction in a lattice model with strong on-site disorder. 
The salient results of these studies show the formation of SPs \cite{Ghosal,Dubi2}, a large spectral gap \cite{Ghosal}, the pinning
of vortices at the regions \cite{Dubi2} where PA were small in the absence of magnetic field, magnetic-field-driven SIT \cite{Dubi} for
the loss of percolation coherence between SPs and diminishing of local correlation at the vortices because of PFs \cite{Erez}, and disorder-driven SIT 
due to strong PFs into a phase of disordered performed pairs \cite{Nandini,Seibold}. 
While all these studies \cite{Dubi,Nandini,Kamlapure,Dubi2,Erez,Seibold} are made for large $U (>3)$, the predicted shell-like effect here is 
for smaller values of $U$ where little changes in local densities cause huge changes in PAs. Therfore, it will not be surprising if some of the qualitative
physics studied earlier for SIT change due to shell-like effect for smaller $U$, since the strong PFs may play a role in the local density fluctuations.


In conclusion, we have found compelling numerical evidence for the shell-like effect to occur in a strongly disordered superconductor because
of the emergent inhomogeneity \cite{Pratap_new} in the form of superconducting puddles. The phase fluctuations and the long-rage Coulomb interaction between
electrons have been ignored in our calculations, but these will not have any qualitative effect since the presence of superconducting puddles is the 
key to our study. By tuning electron density, the quasiparticle excitation gap can be appreciably changed.

We are grateful to P. Raychaudhuri for discussions.


\begin{thebibliography}{99}  

\bibitem{BCS} J. Bardeen, L. N. Cooper, and J. R. Schrieffer, Phys. Rev. {\bf 108}, 1175 (1957).
 \bibitem{Altshuler} A. M. Garcia-Garcia, J. D. Urbina, E. A. Yuzbashyan, K. Richter, and B. L. Altshuler, Phys. Rev. Lett. {\bf 100}, 187001 (2008);
                         Phys. Rev. B {\bf 83}, 014510 (2011).
\bibitem{Ovchinnikov} V. Z. Kresin and Y. N. Ovchinnikov, Phys. Rev. B {\bf 74}, 024514 (2006).
\bibitem{Olofsson} H. Olofsson, S. Aberg, and P. Leboeuf, Phys. Rev. Lett. {\bf 100}, 037005 (2008).
\bibitem{Bose} S. Bose, A. M. Garcia-Garcia, M. M. Ugeda, J. Urbina, C. Michaelis, I. Brihuega, and K. Kern, Nat. Mat. {\bf 9}, 550 (2010). 
\bibitem{Goldman} A. M. Goldman, and N. Markovic, Phys.  Today {\bf 51}, No. 11, 39 (1998).
\bibitem{Review1} V. F. Gantmakher and V. T. Dolgopolov, Usp. Fiz. Nauk {\bf 180}, 3 (2010) [Phys. Usp. {\bf 53}, 1 (2010)].
\bibitem{Anderson_loc} P. W. Anderson, Phys. Rev. {\bf 109}, 1492 (1958).
\bibitem{Ghosal} A. Ghosal, M. Randeria, and N. Trivedi, Phys. Rev. Lett. {\bf 81}, 3940 (1998); Phys. Rev. B {\bf 65}, 014501 (2001).
\bibitem{Dubi} Y. Dubi, Y. Meir, and Y. Avishai, Nature (London) {\bf 449}, 876 (2007).
\bibitem{Sacepe} B. Sacepe, C. Chapelier, T. I. Baturina, V. M. Vinokur, M. R. Baklanov, and M. Sanquer, Phys. Rev. Lett. {\bf 101}, 157006 (2008).
\bibitem{Goldman3} K. H. S. B. Tan K. A. Parendo, and A. M. Goldman, Phys. Rev. B {\bf 78}, 014506 (2008). 
\bibitem{Baturina2} B. Sacepe, C. Chapelier, T. I. Baturina, V. M. Vinokur,  M. R. Baklanov, and M. Sanquer, Nature Communications {\bf 1}, 140 (2010).
\bibitem{Pratap1} M. Mondal, A. Kamlapure, M. Chand, G. Saraswat, S. Kumar, J.  Jesudasan, L. Benfatto, V. Tripathi, 
                 and P. Raychaudhuri, Phys. Rev. Lett. {\bf 106}, 047001 (2011).
\bibitem{Feigelman} M. V. Feigel'man, L. B. Ioffe, V. E. Kravtsov, and E. A. Yuzbashyan, Phys. Rev. Lett. {\bf 98}, 027001 (2007).
\bibitem{Ioffe} M. V. Feigel'man, L. B. Ioffe, V. E. Kravtsov, And E. Cuevas, Ann. Phys. (N.Y.) {\bf 325}, 1390 (2010).
\bibitem{Anderson} P. W. Anderson, J. Phys. Chem. Solids. {\bf 11}, 26 (1959).
\bibitem{AG} A. A. Abrikosov and L. P. Gor'kov, Zh. Eksp. Teor. Fiz. {\bf 35}, 1558 (1958) [Sov. Phys. JETP {\bf 8}, 1090 (1959)].
\bibitem{Ma} M. Ma and P. A. Lee, Phys. Rev. B {\bf 32}, 5658 (1985).
\bibitem{Kapitulnik} A. Kapitulnik and G. Kotliar,  Phys. Rev. Lett. {\bf 54}, 473 (1985).
\bibitem{Fisher} M. P. A. Fisher, Phys. Rev. Lett. {\bf 65}, 923 (1990).
\bibitem{Nandini} K. Bouadim, Y. L. Loh, M. Randeria, and N. Trivedi, Nature Phys. {\bf 7}, 884 (2011).
\bibitem{Goldman2} K. A. Parando,  K. H. S. B. Tan, A. Bhattacharya, M. Eblen-Zayas, N. E. Staley, and A. M. Goldman, Phys. Rev. Lett. {\bf 94}, 197004 (2005).
\bibitem{DeGennes} P. G. De-Gennes,  {\it Superconductivity of Metals and Alloys} (W. A. Benjamin, New York, 1966).
\bibitem{note2} Since the SIT takes place around the critical \cite{Nandini} disorder strength $V_c \sim 1.6$, our chosen parameters are
    either near SIT region or in the deep insulating region.

\bibitem{note} Although the disorder potential remains unchanged, $n_i$ changes
              with $\langle n \rangle$ in such a way that $n_i/\langle n \rangle $ remains independent of $\langle n \rangle$ 
              [see supplemental material (SM)]. 

\bibitem{note1}  $\Delta_{{\rm av}}$ is monotonic for $V \lesssim 1$. The oscillation increases with the increase of $V$ for a moderate
  value of $U$. The normalized oscillation also increases on lowering $U$ for a moderate value of $V$. (See SM). 

\bibitem{Kumar} S. Kumar and P. B. Chakraborty, arXiv:1302.1967v1.
\bibitem{note3} Our analysis is based on the mean field numerical study in small systems which restrict us to considering small values of $U$, while the
real thermodynamic systems may correspond to weak coupling. Now the question arises whether or not the theory presented here will be valid for
the weak coupling superconductors. The predictions of the formation  of superconducting islands in small systems is recently confirmed in a
weak coupling semianalytic theory \cite{Ioffe} for thermodynamic systems and in experiments \cite{Sacepe}. 
The fact that the present theory is applicable for the disorder
range where SPs are formed and that the phenomenon associated with the shell effect \cite{Bose,Altshuler} 
is related to the SPs and gets enhanced on lowering $U$ (see SM), predicts that the effect will hopefully
be applicable to weak coupling thermodynamically large superconductors.


\bibitem{Kamlapure} G. Lemarie, A. Kamlapure, D. Bucheli, L. Benfatto, J. Lorenzana, G. Seibold, S. C. Ganguli, P. Raychaudhuri,
                   and C. Castellani, Phys. Rev. B {\bf 87}, 184509 (2013).
\bibitem{Dubi2} Y. Dubi, Y. Meir, and Y. Avishai, Phys. Rev. B {\bf 78}, 024502 (2008).
\bibitem{Seibold} G. Seibold, L. Benfatto, C. Castellani, and J. Lorenzana, Phys. Rev. Lett. {\bf 108}, 207004 (2012).
\bibitem{Erez} A. Erez and Y. Meir, EPL {\bf 91}, 47003 (2010).
\bibitem{Pratap_new} A. Kamlapure, T. Das, S. C. Ganguli, J. B. Parmar, S. Bhattacharyya, and P. Raychaudhuri,
                     Sci. Rep. {\bf 3}, 1 (2013). 
 



















\end{thebibliography}
\end{document}


\title{Supplemental Material for ``Amplitudle Fluctuations Driven by the Density of Electron Pairs within Nanosize Granular Structuters 
inside Strongly  Disordered Superconductors: Evidence for a Shell-Like Effect''}
\author{Sanjib Ghosh and Sudhansu S. Mandal}
\affiliation{Department of Theoretical Physics, Indian Association for the Cultivation of Science,
           Kolkata 700 032, India}

 

\maketitle

\section{Density Profile}

\begin{figure}[t]
\centering
\includegraphics[scale=0.7]{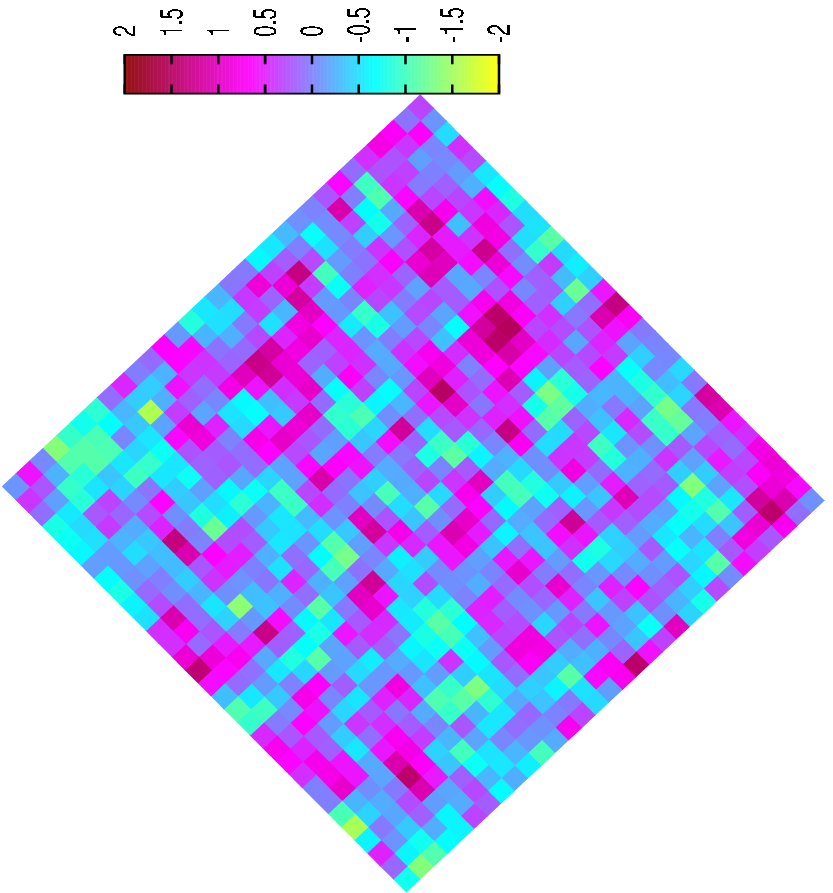}
\caption{Disorder profile for V=2 that has been used for calculting Fig.~1 and most of the results shown in the paper. }
\label{FigS1}
\end{figure}

\begin{figure}[h]
\centering
\includegraphics[scale=0.6]{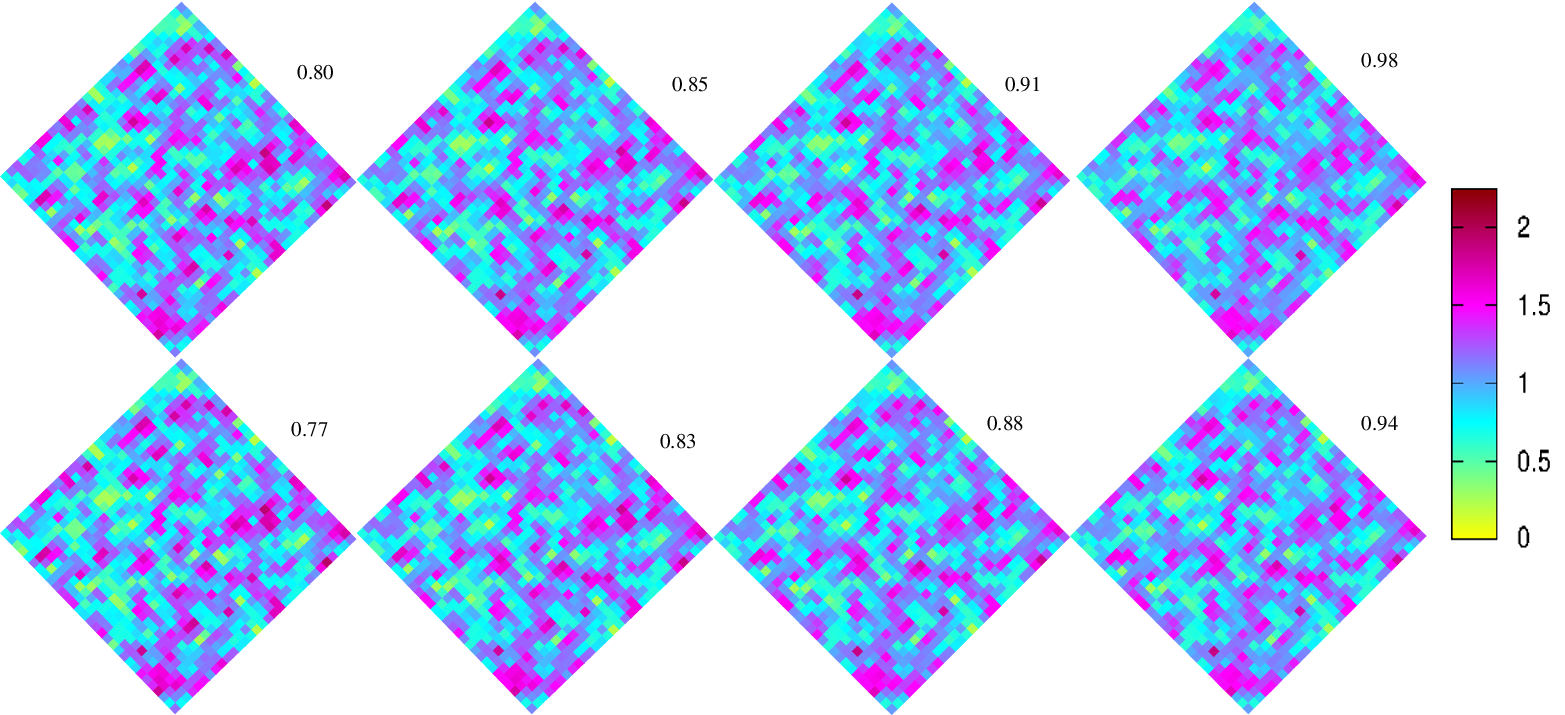}
\caption{Color-scale plot of $n_i/\langle n \rangle$ at different sites in a $32\times 32$ square lattice with $U=1.5$, $V=2$, and
  a fixed realization of disorder which has been used for generating data shown in Fig.~1. Number adjacent to each panel represents
 the corresponding value of $\langle n \rangle$. }
\label{FigS2}
\end{figure}

\begin{figure}[h]
\centering
\includegraphics[scale=0.7]{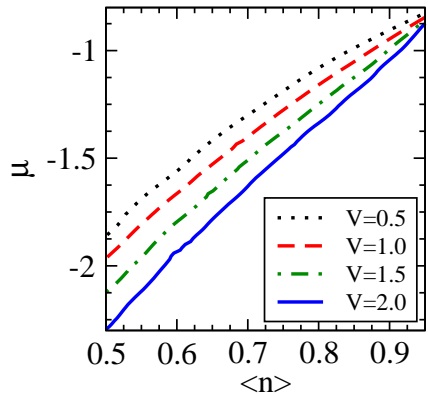}
\caption{The dependence of chemical potential, $\mu$, on $\langle n \rangle$ for $U=1.5$ and different values of $V$.}
\label{FigS3}
\end{figure}

 In Fig.~1, we have shown the profile of $\Delta_i$ at different values of $\langle n \rangle$ with $V=2$, $U=1.5$, and fixed disorder
realization (Fig.~\ref{FigS1}).  In figure \ref{FigS2}, we show the corresponding profiles for $n_i$. We note that although it is expected
that $n_i$ will change as $\langle n \rangle$ changes, the profile for $n_i/\langle n \rangle$ remains alomost unchanged. 
 Figure \ref{FigS3} shows the variation of chemical potential with $\langle n \rangle$ at different values of $V$.

\section{Pair-amplitude Fluctuation}

\begin{figure}[h]
\centering
\includegraphics[scale=0.5]{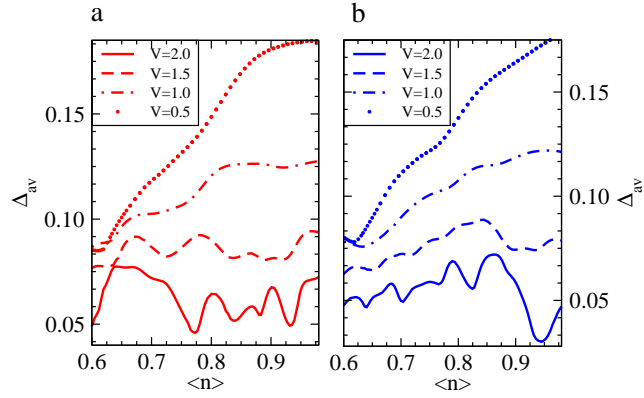}
\caption{Average pair-amplitude, $\Delta_{{\rm av}}$, in two $10\times 10$ regions marked as A and B in the left bottommost panel 
of Fig.~1 versus $\langle n \rangle$ for different values of $V$ at $U=1.5$. Left (right) panel corresponds to region A (B).
 For all values of $\langle n \rangle$, same disorder landscape has been used.
While $\Delta_{{\rm av}}$ is monotonic with $\langle n \rangle$ at smaller values of $V$, it starts oscillating around $V>1$.
The oscillation increases with the increase of $V$. }
\label{FigS4}
\end{figure}

\begin{figure}[h] 
\centering
\includegraphics[scale=0.6]{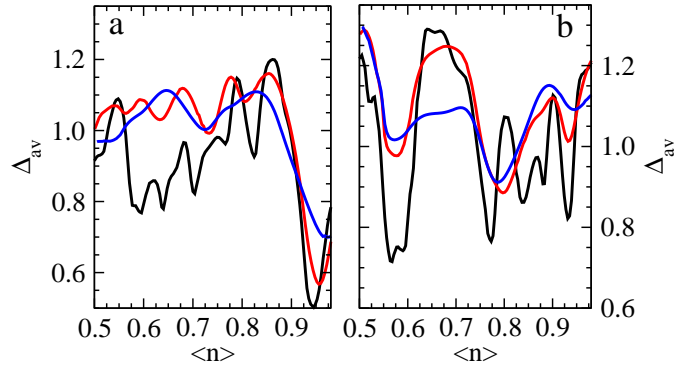} 
\caption{Same as in Fig.~\ref{FigS4} but for a fixed $V=2$ and different values of $U$: 1.5 (black), 2.0 (red), 3.0 (blue). 
     $\Delta_{{\rm av}}$ have been scaled with scaling factors: 0.06 ($U=1.5$), 0.13 ($U=2.0$), 0.31 ($U=3.0$).
Left (right) panel corresponds to region B (A) in Fig.~1. The
oscillation of $\Delta_{{\rm av}}$ with $\langle n \rangle$ increases with decreasing $U$.}
\label{FigS5}
\end{figure}

 In Fig.~3, we have shown that $\Delta_{{\rm av}}$ calculated for two selected regions of size 10$\times 10$ each oscillates with $\langle n\rangle $
with fixed $U$, $V$, and realization of disorder. In Fig.~\ref{FigS4}, we show the variation of $\Delta_{{\rm av}}$ with $\langle n \rangle$
at these two regions with fixed $U$ and realization of disorder for diffrent values of $V$. The oscillation begins when $V>1$, and it increases with
the increase of $V$. In Fig.~\ref{FigS5}, we show $\Delta_{{\rm av}}$ for fixed $V$ and different values of $U$.  The oscillation in $\Delta_{{\rm av}}$
increases with decreasing $U$.

